\documentclass[baaa]{baaa}

\usepackage[pdftex]{hyperref}
\usepackage{subfigure}
\usepackage{natbib}
\usepackage{helvet}
\usepackage[font=small]{caption}

\begin{document}


\journalvol{57}
\journalyear{2014}
\journaleditors{A.C. Rovero, C. Beaugé, L.J. Pellizza \& M. Lares}


\contriblanguage{0}


\contribtype{3}

\thematicarea{6}

\title{Impacto ambiental de los Remanentes de Supernova}


\titlerunning{Impacto ambiental de RSN}


\author{G.M. Dubner\inst{1}}
\authorrunning{Dubner}
\contact{gdubner@iafe.uba.ar}

\institute{Instituto de Astronomía y Física del Espacio (IAFE, CONICET-Universidad de Buenos Aires)
}


\resumen{Al explotar una supernova (SN) inyecta en forma casi instantánea unos 10$^{51}$ ergios de energía térmica y mecánica en una pequeña región del espacio, originando así la formación de poderosas ondas de choque que se propagan a través del medio interestelar a velocidades de varios miles de km/seg. Estas ondas barren, comprimen y calientan la materia que encuentran formando los remanentes de supernovas, cuya evolución a lo largo de miles de años transforma para siempre, irreversiblemente, no sólo las propiedades físicas sino también la química de una vasta región del espacio que puede abarcar centenares de parsecs. En esta contribución se analiza brevemente el impacto que tienen estas explosiones discutiendo, sobre la base de resultados teóricos y observacionales recientes, la relevancia de  algunos de los fenómenos comúnmente asociados con SN y sus remanentes.}

\abstract{
The explosion of a supernovae (SN)  represents the sudden injection of about $ 10^{51} $ ergs of thermal and mechanical energy in a small region of space, causing the formation of powerful shock waves that propagate through the interstellar medium at  speeds of several thousands of km/s. These waves sweep, compress and heat the interstellar material that they encounter,  forming the supernova remnants. Their evolution  over thousands of years change forever, irreversibly, not only the physical but also the chemical properties of a vast region of space that can span hundreds of parsecs. This contribution briefly analyzes the impact of these explosions, discussing the relevance of some phenomena usually associated with SNe and their remnants in the  light of recent theoretical and observational results.
}

%
%




\maketitle

\section{Introducción}
\label{S_intro}

\noindent

Al final de sus vidas algunas estrellas pueden sufrir un colapso de origen termonuclear o gravitacional y terminan explotando como supernovas (SN). En un evento de SN, en fracción de segundos puede liberarse una energía equivalente a unos 10\,000 millones de Soles en un punto y entre 5 y 10 M$_\odot$  salen despedidas a un 3 \% de la velocidad de la luz.  En una galaxia como la Vía Láctea explotan unas 2 a 3 SN  por siglo,  en el Universo visible explotan unas 8 SN por segundo, de modo que al cabo de  1 hora aparecen casi 30\,000 SN nuevas en el  Universo.

 Estas explosiones destruyen la estrella original y ge\-ne\-ran episodios de nucleosíntesis explosiva que fabrican elementos radiactivos, lo que las hace visibles a grandes distancias en el Universo. Desde la perspectiva de las galaxias, representan la inyección súbita de unos 10$^{51}$ ergios de energía térmica y mecánica en una pequeña región del espacio, originando la formación de poderosas ondas de choque que se propagan a través del medio interestelar a velocidades de varios miles de km/seg. Estas ondas barren, comprimen y calientan la materia que encuentran, formando así los remanentes de supernovas (RSN). Además, al explotar liberan al espacio los elementos atómicos sintetizados en el interior de las es\-tre\-llas durante toda su vida, modificando el estado físico y químico  de una inmensa región del espacio que puede abarcar centenares de parsecs. 

Los RSN son una de las  principales fuentes de energía y transformación de las galaxias y su impacto ambiental es enorme, originando  variados procesos físicos que son de gran interés en la astrofísica actual: desde la a\-ce\-le\-ra\-ción de rayos cósmicos hasta la química prebiótica en un planeta, incluyendo nucleosíntesis explosiva, formación y destrucción de moléculas y polvo,  evolución estelar, formación de los objetos más densos del Universo, etc. La investigación de los RSN es entonces necesariamente una confluencia multidisciplinaria que incluye astrofísica, astroquímica, física de altas energías, cosmología, física de partículas, física del plasma, astrobiología y hasta arqueoastronomía. A su vez, por las características de las emisiones que se originan durante la evolución de un RSN, éstos pueden radiar a través de todo el espectro electromagnético, desde ondas de radio hasta rayos $\gamma$ ultra-energéticos, requiriendo de diferentes instrumentos terrestres y espaciales para una investigación completa.

 En los últimos años ha habido un rápido progreso en el conocimiento tanto teórico como observacional de los RSN. En esta contribución se analiza brevemente la relevancia de los principales fenómenos que usualmente se asocian con la explosión de una SN y sus remanentes en el espacio, haciendo una revisión  crítica  de los mismos sobre la base de los resultados más recientes publicados en el tema.

\section{Mitos y verdades de los RSN}

Históricamente se ha afirmado que los RSN son agentes de profundos cambios en las galaxias. Analizaremos brevemente algunas de las aseveraciones más frecuentes a la luz de las últimas investigaciones en el tema.

\subsection{Las SN son los principales fabricantes de átomos del Universo}
Luego de  los primeros  minutos del Universo (probablemente no más de 20 minutos) en los que tuvo lugar la nucleosíntesis primordial formando  los nucleones primigenios (H- He–Li)  a partir del plasma de quarks y gluones del Big Bang,  la nucleosíntesis estelar es responsable a través de  procesos sucesivos de fusión nuclear dentro de las estrellas de la generación de los elementos atómicos por lo menos  hasta el $^{56}$Fe. 

Al momento de la explosión una SN no sólo libera todos los núcleos formados en el seno de la estrella progenitora, sino que en brevísimo tiempo se produce lo que se conoce como nucleosíntesis explosiva, dando origen a la formación de más de 60 núcleos diferentes. Más de la mitad de los núcleos de la Tabla Períódica  son fabricados al momento de la explosión por las altas tem\-pe\-ra\-tu\-ras asociadas con el pasaje de la onda explosiva y la gran abundancia de neutrones, a través de tres tipos diferentes de procesos: los  s (por {\it slow}), los r  (por {\it rapid}) y los p (por {\it photodisintegration}),  o sea la  fotodesintegración de núcleos pre-existentes creados por procesos-s, que son responsables de la creación de muchos de los elementos radioactivos, como uranio, torio, etc. La prueba más convincente de nucleosíntesis explosiva en SN se tuvo con la SN 1987A, cuya emisión de líneas en rayos $\gamma$ permitió identificar la presencia de núcleos de  $^{56}$Co y $^{57}$Co.

\begin{figure}[!ht]
  \centering
  \includegraphics[width=0.45\textwidth]{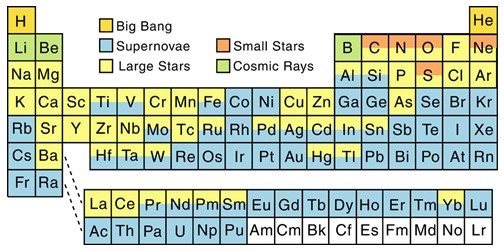}
  \caption{Importante rol de las SN en la producción y liberación de elementos atómicos al espacio (tomada de http://www4.nau.edu/meteorite/Meteorite/Book-GlossaryN.html)  }
  \label{tablaperiodica}
\end{figure}

Estos procesos  convierten a las SN en la fuente dominante de elementos pesados en el Universo. Son la única fuente conocida de elementos alfa (O, Ne, Mg, Si, S, Ar y Ca) y elementos del grupo del Fe (Fe y Ni). La liberación y posterior condensación de estos núcleos, junto con la pérdida de masa de las estrellas gigantes, tiene un rol especial en la creación de polvo interestelar. La Figura \ref{tablaperiodica} ilustra el importante rol de las explosiones estelares como proveedoras de elementos atómicos en el Universo.  Una buena síntesis de la historia de la materia desde el Big Bang hasta el presente puede encontrarse en \citet{arnett1996}.

\subsection{Son los grandes modificadores de la dinámica de las galaxias}
\begin{figure}[!ht]
  \centering
  \includegraphics[width=0.45\textwidth]{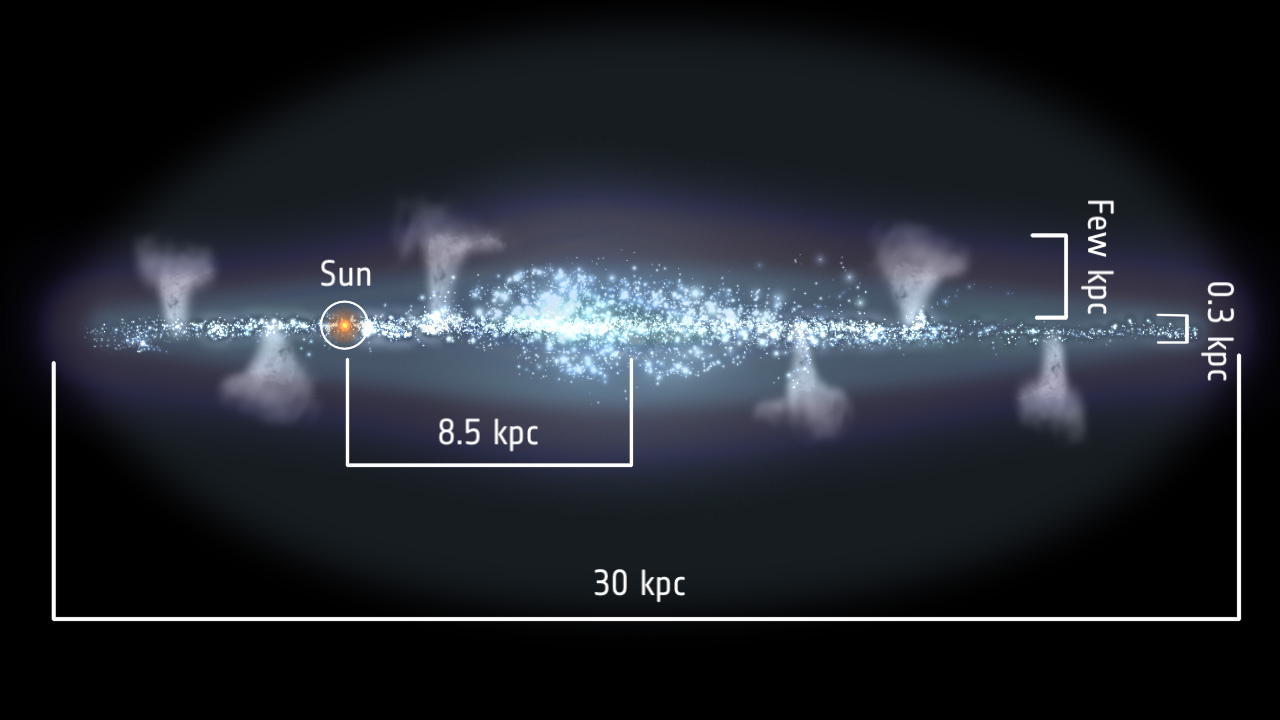}
  \caption{Imagen artística representando  fuentes galácticas de gas caliente emitiendo en rayos X en el halo de la Vía Láctea, creadas por la turbulencia inyectada por explosiones de SN en el medio interestelar. Crédito ESA. }
  \label{galacticfountains}
\end{figure}

Las explosiones de SN son, efectivamente,  la fuente más importante de energía mecánica y térmica en el medio interestelar. Los vientos estelares aportan cantidades similares de energía pero lo hacen gradualmente a lo largo de un millón de años. Las SN lo hacen en segundos o días como máximo. Son la principal fuente de calentamiento del gas difuso. En las galaxias pueden formar túneles, chimeneas o plumas sobresaliendo el disco (Fig. \ref{galacticfountains}).
 
En particular la interacción de los RSN que provienen de colapso gravitacional con su complejo entorno (burbujas de viento, paredes de cavidades y nubes moleculares pertenecientes al complejo donde probablemente nació la estrella precursora) es la principal fuente de transferencia de masa y energía entre estrellas y medio gaseoso en las galaxias.
 Las ondas de choque de un RSN pueden comprimir, calentar, excitar, ionizar y  disociar moléculas, como así también contribuir a la formación de nuevas especies moleculares. En nuestra Galaxia, aproximadamente unos 70 RSNs podrían estar interactuando físicamente con nubes moleculares \citep{ chen2014}.  Estas interacciones pueden  excitar máseres de OH (que emiten en 1720 MHz)  y, como se verá más adelante, acelerar partículas hasta energías que se vuelven emisoras de rayos $\gamma$ a través de interacciones hadrónicas. Tal como resume \citet{slane2014}, estas interacciones juegan un papel  muy importante en la comprensión de la naturaleza física de los RSN.

\subsection{Son la fuente de los rayos cósmicos de origen galáctico}
Desde 1953 cuando Shklovskii especuló que los rayos cósmicos (RC) podrían ser acelerados en las nebulosas de los RSN, es un hecho aceptado por consenso general que los RSN son los candidatos favoritos como fuente de rayos cósmicos galácticos, por lo menos con energías hasta la ``rodilla''  del espectro ($\sim 10^{15}$ eV).  Históricamente esta afirmación se ha basado en tres argumentos: (1) las SN galácticas son suficientes como fuente de energía: la inyección  de energía requerida para explicar la energía de los RC observados en Tierra es aproximadamente de un 10 a un 30\% de la energía cinética entregada al medio interestelar (MIE) por explosiones de SN; (2) la observación de radiación sincrotrónica en ondas de radio en los RSN implica que hay electrones acelerados a energías por lo menos hasta los GeV, y desde que se detectó radiación sincrotrónica en rayos X en los RSN: RXJ1713.7-3946, Vela Jr, RCW86 y SN1006, se sabe que los frentes de choque de RSN pueden acelerar electrones por lo menos hasta los TeV; no hay evidencia fuerte que se aceleren también iones, pero hay modelos viables; (3) un modelo simple de aceleración difusiva en choques (DSA por su sigla en inglés) en el límite de partícula de prueba, predice un espectro con la ley de potencia correcta. Si bien predice -2.0 y en Tierra tienen un exponente -2.7, la discrepancia puede explicarse con argumentos de tiempo de confinamiento de las partículas. Sin embargo, tras más de 100 años del descubrimiento de los RC todavía nos preguntamos acerca de su origen exacto y hasta su naturaleza misma (sobre todo para los de muy alta energía).  Y cuanto más avanzan las investigaciones, más dudas surgen. 

El principal desafío es identificar el origen de los iones acelerados \citep{butt2009}.
 La manera más directa de encontrar sitios donde se aceleren partículas es buscar emisión $\gamma$ coincidente o cerca de fuentes sospechadas de ser aceleradores. Si un objeto acelera RC, se espera que en la vecindad haya una superpoblación de partículas recién aceleradas. Esa nube de partículas energéticas puede  interactuar con la materia y la radiación ambiente y producir rayos $\gamma$ que vemos desde Tierra. El problema es que tanto iones como electrones en los rayos cósmicos producen un efecto parecido, y es extremadamente difícil determinar qué tipo de partículas produjo los rayos $\gamma$. Hay algunas evidencias de aceleración de iones de rayos cósmicos en RSN, por ejemplo la separación entre el frente de choque y la discontinuidad de contacto (o choque reverso) aparece considerablemente reducida en el RSN de Tycho  y eso se explica porque se gastó una cantidad considerable de energía en acelerar iones de rayos cósmicos  \citep{warren2005}.  Igualmente, de esa observación no se puede concluir que los RSN de nuestra Galaxia sean la fuente principal de RC.

\begin{figure*}[!ht]
  \centering
  \includegraphics[width=1\textwidth]{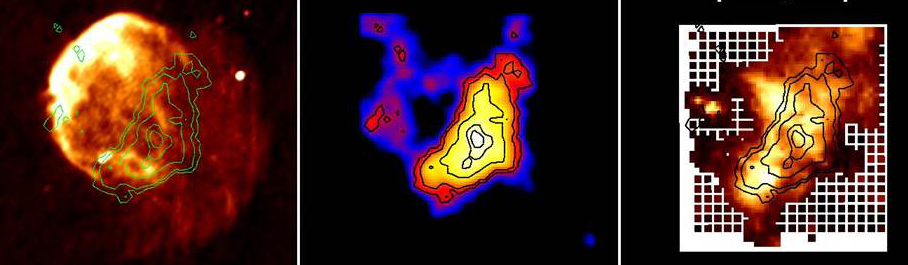}
  \caption{Producción de rayos $\gamma$ a través de interacción del RSN IC443 con una nube molecular. {\it  Izquierda:} el RSN IC443 en radio a 330 MHZ \citep{castelletti2011};  {\it Centro:} emisión gamma en TeV detectada por VERITAS \citep{acciari2009}; {\it Derecha:} emisión en CO \citep{zhang2010}. En todos los casos los contornos muestran la localización de la emisión $\gamma$.}
  \label{ic443}
\end{figure*}

Hay otras  evidencias observacionales de aceleración de RC en algunos RSN viejos, como por ejemplo el RSN IC443 (Fig. \ref{ic443}),  cuya emisión $\gamma$ de origen hadrónico fue detectada con VERITAS \citep{acciari2009} y recientemente \citet{ackermann2013} detectaron con Fermi LAT las evidencias características de protones acelerados como consecuencia del encuentro de  los frentes de choque de los RSN IC443 y W44  con nubes moleculares del entorno.

¿Cuál es el problema con la descripción standard? Que los tres argumentos más fuertes: que los RSN son las únicas fuentes conocidas en la Galaxia con la energía  suficiente, que el espectro es más o menos el correcto y que los RSN aceleran electrones, siguen siendo indirectos. Habría otras fuentes en la Galaxia, por ejemplo la rotación galáctica (con reconexión magnética y ondas de densidad), {\it jets} de estrellas de neutrones y agujeros negros acretando, pulsares, nebulosas de viento de pulsar, etc.
Una propuesta con aceptación creciente es no tomar RSN aislados, sino aglomeraciones de ellos junto con estrellas masivas formando superburbujas \citep{binns2007}. 

Algunos aspectos pendientes de solución son:  los electrones energéticos ($\geq$ 100 GeV) pierden la energía mucho más rápido que los iones. Entonces si se los ve en Tierra, sus fuentes deben estar confinadas a unos pocos kpc del Sol, mientras que los iones pueden estar mucho más lejos. Asi que las fuentes de electrones en RC pueden ser diferentes que las fuentes de iones.
Además, las anomalías isotópicas observadas en RC no se corresponden con las anomalías observadas en RSN y sí en superburbujas. Y otro aspecto difícil de explicar es la muy leve anisotropía de los RC, que no se condice con la localización preferencial de los RSN.
En resumen: probablemente los mecanismos de aceleración de iones en RSN son más complejos que los propuestos al presente, y aún si los RSN son los aceleradores de iones, no hay pruebas suficientes que RSN aislados sean la principal fuente de iones en RC.

\subsection{Las SN y sus remanentes son fábrica de moléculas y de polvo}
Un tema astrofísico de gran importancia es cuál es el origen del polvo en las galaxias, sobre todo en el Universo temprano.
Frecuentemente se menciona que las SN son agentes que aportan polvo al MIE. Efectivamente, en una explosión de SN se inyectan unas 3 M$_\odot$ de elementos pesados. De éstos $\sim$ 1\% se espera que esté en forma de polvo; si hay unas 3 SN cada 100 años (en galaxias tipo Vía Láctea), la tasa de formación de polvo por SN es $\sim$ 0.01 M$_\odot$/año, comparable a la acción de estrellas gigantes rojas. La detección de 0.1 a 0.5 M$_\odot$ de polvo en RSN cercanos sugiere que se forma polvo en los primeros años tras la explosión.  Se calcula que un grano de polvo vive $\sim$ 100 años y durante su vida es reciclado al menos 10 veces entre nube y medio internube, con una importante influencia de las SN en el reciclado \citep{micelotta2013}.

 La mayoría de los elementos refractarios  se producen durante explosiones de SN, pero no está claro cómo y dónde se condensan y crecen los granos de polvo, y cómo evitan su destrucción en un ambiente hostil como el entorno de SN y regiones de formación estelar. \citet{gall2014} reportaron recientemente la formación rápida (entre los 40 y 240 días tras la explosión)  de granos de polvo grandes en la SN2010j. La primera formación de polvo ocurre en una cáscara densa y fría detrás del frente de choque (en la eyecta es imposible porque está muy caliente). A tiempos tardíos (500 a 900 días después de la explosión) se observó un crecimiento acelerado de la masa de polvo, marcando la transición del medio circumestelar al eyecta. En ese momento el polvo se está formando en trozos de eyecta que viajan a unos 7500 km/s. 

Observaciones muy recientes realizadas con ALMA  de SN1987A \citep{indebetouw2014}  en 450 $\mu$m, 870 $\mu$m, 1.4 mm y 2.8 mm  detectaron emisión de la mayor masa de polvo medida en un RSN (más de 0.2 M$_\odot$). Por primera vez se demostró sin ambigüedades que el polvo se formó en la eyecta, los restos fríos del núcleo de la estrella que explotó. La emisión de polvo está concentrada en el centro del remanente, asi que el polvo todavía no fue afectado por los choques. Si sobrevive una buena fracción de este polvo, y si SN1987A es típica, entonces se probaría que las SN son productores de polvo muy importantes en las galaxias.

Por otro lado hay que tener en cuenta que los RSN viejos son agentes muy importantes de destrucción de polvo interestelar vía {\it ``grain sputtering''}  detrás del choque.  Los choques rápidos (v $\sim$ 300 km/s) destruyen los granos pequeños, mientras que  los choques lentos (v$\leq$ 200 km/s) los  vaporizan.

En conclusión, las SN jóvenes son muy eficientes para sintetizar polvo (en la fase temprana de expansión libre), pero al mismo tiempo representan el mayor agente responsable por su destrucción, durante la fase subsecuente del remanente.

Con respecto a la formación de moléculas, el descubrimiento reciente de líneas rotacionales de CO y SiO en la eyecta de SN1987A con ALMA \citep{kamenetzky2013}, fue la primera vez que  mostró tal emisión en un RSN. Detectaron unas 0.01 M$_\odot$ de CO confinado en un volumen esférico que se está expandiendo a ~2000 km/s.  Antes, a los 192 días de la explosión se había observado CO, pero luego de los 600 días desapareció, cuando el gas se volvió muy frío para excitar transiciones vibracionales. Lo que se observa ahora es CO recién formado. En otras 8 supernovas se observó CO en los primeros años tras la explosión, así como  en el RSN joven Cas A \citep{rho2012}.

\subsection{Los RSN pueden originar el nacimiento de estrellas nuevas}

\begin{figure*}[!ht]

  \includegraphics[width=\textwidth]{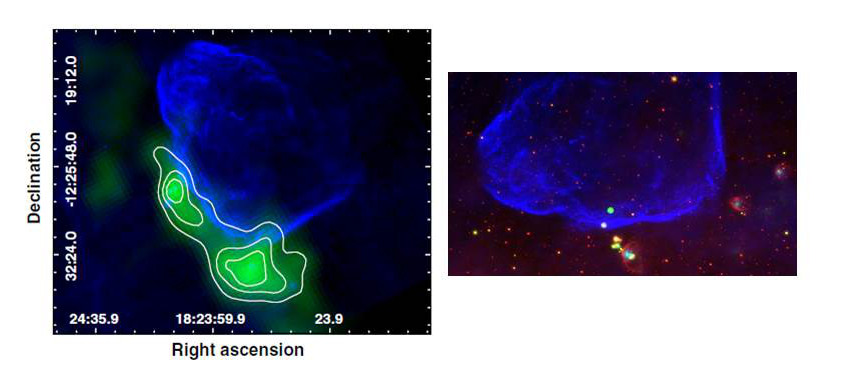}
  \caption{{\it Izquierda:} El RSN G18.8+0.3 en ondas de radio (azul) está rodeado por nubes moleculares (verde) \citep{dubner1999} ; {\it Derecha:} Imagen (mostrada en coordenadas galácticas) de la emisión en radio combinada con datos {\it Spitzer} en infrarrojo en  8$\mu$m y en 24 $\mu$m   que muestra varias  regiones de formación estelar embebidas en la nube (pequeñas burbujas rojas cerca del borde inferior y derecho del RSN). Estudios detallados de las mismas mostraron que la formación estelar no está  relacionada con el RSN. \citep{paron2012} }
  \label{g18}
\end{figure*}

Se ha sugerido a menudo que los frentes de choque de RSN pueden desencadenar la formación de es\-tre\-llas nuevas.  Es cierto que en los alrededores de RSN muy frecuentemente se ven regiones de formación estelar. Eso es lógico porque suelen convivir en el mismo vecindario, ya que los  RSN que provienen de colapso gravitacional de estrellas de alta masa viven pocos años y no se apartan demasiado de la nube molecular madre en la que nacieron. Pero eso no implica que necesariamente estén causalmente conectados.
Una región de formación estelar es una mezcla turbulenta de nubes de gas atómico, gas molecular y polvo, que interactúan bajo la influencia de campos magnéticos, campos de radiación, turbulencia y sobre todo gravedad (puede complicarse aún un poco más con rayos cósmicos, campos de radiación externos y ondas de choque). Se requiere un aumento de presión localizado para iniciar la formación de estrellas nuevas. Los cálculos teóricos muestran que choques a $\sim$ 20 a 45 km/s  pueden comprimir nubes moleculares y desencadenar la formación de grumos autogravitantes que eventualmente pueden terminar siendo estrellas. Pero choques más rápidos aumentan la turbulencia y hasta pueden destruir las nubes moleculares, y los RSN tienen la mayor parte de sus vidas velocidades de varios centenares de km/s y temperaturas mayores a 100 K. En la Figura \ref{g18} se muestra el caso del RSN Galáctico G18.8+0.3, cuya interacción fisica con una nube molecular densa y la existencia de regiones de formación estelar activa embebidas en dicha nube sugerían una posible conexión genética. Un estudio detallado realizado por \citet{paron2012} mostró que el frente de choque del RSN aún no alcanzó a pe\-ne\-trar la nube y posiblemente el RSN y los objetos estelares jóvenes sean coetáneos. 

\citet{desai2010} realizaron un estudio muy completo de todos los RSN identificados en la Nube Mayor de Magallanes, investigando regiones de formación estelar en las inmediaciones. La muestra incluyó los 45 RSN identificados, y se buscaron objetos estelares jóvenes (YSOs por su sigla en inglés)  en yuxtaposición con RSN.  Una vez identificados RSN, nube molecular (NM) e YSO, investigaron  la ubicación de frentes de ionización y evidencias de RSN interactuando con las nubes moleculares. De 45 RSN, encontraron: 7 (RSN + YSO + NM), 3 (RSN + YSO), 8 (RSN + NM). Entre  los 10 SNR que tenían YSOs en las cercanías, 2 están en regiones con formación estelar activa, de modo que es difícil decidir la influencia efectiva de los RSN y  en 4 casos los YSO están claramente fuera del borde de los RSN o la asociación es incierta. En los 4 casos restantes los YSO aparecen proyectados sobre el borde del RSN. Mirando en más detalle se encuentra que en 2 casos los YSO están asociados con regiones HII vecinas. En los últimos 2 casos, los tiempos son incompatibles (el tiempo desde que el RSN empezó a interactuar con la nube natal de los YSO es mucho menor que las escalas de tiempo de los YSO). Resultado: en la Nube Mayor de Magallanes los RSN no han actuado como formadores de estrellas nuevas en ningún caso.

 Cuando los RSN envejezcan y sus velocidades hayan disminuido por debajo de los ~45 km/s (y los choques no sean destructivos) los YSO que se observan ahora ya van a ser estrellas más evolucionadas y sus frentes de ionización podrán a su vez estar desencadenando la formación de estrellas nuevas en el mismo vecindario, haciendo muy compleja la identificación de los distintos mecanismos en acción. Lo que sí se puede concluir  es que en todos los casos los RSN alteran las propiedades físicas de los YSO, ya que ellos interceptan elementos pesados y elementos radioactivos, los cuales serán incorporados en las envolturas circumestelares y discos, tal como sucedió en la nebulosa solar.

\subsection{Las SN pueden originar los objetos más compactos del Universo}
En 1934, a sólo 2 años del descubrimiento del neutrón,   Baade y Zwicky   publicaron {\it ``Con toda reserva podemos adelantar la impresión de que una supernova representa la transición de una estrella ordinaria a una estrella de neutrones, consistente principalmente de neutrones. Tal estrella puede ser de muy pequeño radio y extremadamente alta densidad''} \citep{baade1934}. Desde entonces una gran cantidad de trabajos fueron perfeccionando el conocimiento de los mecanismos de colapso que llevan a la formación de estas estrellas y la ecuación de estado de la materia neutrónica que alcanza densidades del orden de $10^{11}$ g cm$^{-3}$. Sin embargo a la fecha estos son aún temas sujetos a intenso debate, y más aún cuando el objeto resultante es un agujero negro.

 Se espera que aproximadamente un 85\% de los RSN, los que provienen de SN de tipo Ib, Ic y II, dejen un objeto compacto. El catálogo más completo y actualizado de RSN\footnote{http://www.mrao.cam.ac.uk/surveys/snrs/} \citep{green2014} lista un total de 294 RSN en nuestra Galaxia.  Desde la última versión (de 2009) se agregaron 21 nuevos remanentes descubiertos y se eliminó 1 que fue reclasificado como región HII.  Por otra parte, el catálogo más completo de pulsares  en su versión 2014\footnote{http://www.atnf.csiro.au/people/pulsar/psrcat/}, lista 2328 pulsares, o sea estrellas de neutrones pulsantes con el haz dirigido hacia Tierra. A ellos se debe agregar toda la población de pulsares cuyo haz no apunta a la Tierra y/o no pueden ser detectados como objetos pulsantes en radio u otras bandas electromagnéticas, los objetos centrales compactos (CCOs), los {\it Anomalous X-ray pulsars (AXPs){\rm  , los} Soft-Gamma Repeaters (SGRs)}, y una posible población de  objetos compactos  que no pulsan. 
Así que hay por lo menos un orden de magnitud de diferencia entre el número de RSN y el número de objetos compactos detectados en nuestra Galaxia. ¿Cómo se cierra esa discordancia? 

Algunas explicaciones propuestas son: las estrellas de neutrones viven mucho más que los RSN. El tiempo de vida de un pulsar típico es de unos 10$^{6-7}$ años, lo cual excede largamente el tiempo de vida de un RSN (del orden de los 10$^{4-5}$ años). También los pulsares escapan  (velocidades transversales de entre 100 y 400 km/s) y no hay manera de asociarlos con su  RSN original. Sin embargo aún siguen abiertas opciones como mecanismos explosivos que generen la formación de objetos compactos sin formar una nebulosa como RSN.  

\subsection {¿ Y cuál sería la impacto de una explosión de SN sobre la ecología terrestre?}
El flujo de rayos $\gamma$ que se produce al momento de explotar una SN puede inducir una reacción química en la alta atmósfera convirtiendo N molecular en óxidos de N. Tal reacción llevaría a vaciar la capa de ozono, exponiendo la Tierra a la radiación UV nociva del Sol. Una SN de tipo Ib, Ic o II, tiene que producirse en algún punto más cerca que 26 años-luz para provocar la destrucción de la capa de ozono. Si se trata en cambio de una SN Ia, aún estando tan lejos como  unos 3000 años-luz, ya sería suficiente para destruirla.

Se encontraron niveles elevados de nitratos en el hielo antártico que coincidirían con los eventos de SN1006 (que explotó a 7200 años-luz) y de la Nebulosa del Cangrejo que explotó en el año 1054 a 6300 años-luz. Posiblemente los rayos $\gamma$ de esas explosiones alcanzaron la Tierra y aumentaron los niveles de óxidos de N, que quedaron atrapados en los hielos.

 \section{Conclusiones}

Los restos de las explosiones de supernova son uno de los objetos más fascinantes del Universo,  una poderosa  fuente de energía muy localizada que puede generar una serie de fenómenos que los convierte en un laboratorio ideal para investigar la física de los extremos. Usualmente se invocan explosiones de SN y la expansión de sus frentes de choque cada vez que se requieren compresiones poderosas, aceleradores de partículas, grandes aportes de energía, etc. Sin embargo no todos estos roles están suficientemente probados. Los estudios más recientes confirman, por ejemplo, que los RSN pueden acelerar rayos cósmicos, pero probablemente no pueden formar estrellas nuevas como se ha afirmado por años. El estudio de los RSN está entre los objetivos científicos de todos los instrumentos de frontera, instalados y proyectados, ya que su conocimiento profundo puede servir para explicar desde la edad y forma del Universo hasta la presencia de vida en nuestro planeta. 

\begin{acknowledgement}
\label{thanks}

Agradezco a los organizadores de la 57a. Reunión Anual de la Asociación Argentina de Astronomía (AAA) por haberme invitado a presentar esta revisión y por facilitar  mi participación en la reunión. El trabajo se realizó con el apoyo de los
 subsidios PIP 0736/11 de  CONICET y PICT 0571/2011 de ANPCyT.  GD es miembro de la Carrera del Investigador Científico de CONICET.

\end{acknowledgement}


\bibliographystyle{baaa}
\small
\bibliography{dubner}
 
\end{document}